# A Signal Matrix-Based Local Flaw Detection Framework for Steel Wire Ropes Using Convolutional Neural Networks

Siyu You[1], Leilei Yang[2], Zixu Kuang[1], Huayi Gou[1], Longlong Zhang[3], Zhiliang Liu[2]*
[1] Glasgow College, University of Electronic Science and Technology of China, Chengdu, China
[2] School of Mechanical and Electrical Engineering, University of Electronic Science and Technology of China, Chengdu China
[3] Shinva Medical Instrument Co., Ltd., Zibo, Shandong, China
*Corresponding author: Zhiliang Liu (zhiliang_liu@uestc.edu.cn)

*Abstract*— Steel wire ropes (SWRs) are critical load-bearing components in industrial applications, yet their structural integrity is often compromised by local flaws (LFs). Magnetic Flux Leakage (MFL) is a widely used non-destructive testing method that detects defects by measuring perturbations in magnetic fields. Traditional MFL detection methods suffer from critical limitations: one-dimensional approaches fail to capture spatial relationships across sensor channels, while multi-dimensional image-based techniques introduce interpolation artifacts and computational inefficiencies. This paper proposes a novel detection framework based on signal matrices, directly processing raw multi-channel MFL signals using a specialized Convolutional Neural Network for signal matrix as input (SM-CNN). The architecture incorporates stripe pooling to preserve channel-wise features and symmetric padding to improve boundary defect detection. Our model achieves state-of-the-art performance with 98.74% accuracy and 97.85% recall. Additionally, it demonstrates exceptional computational efficiency, processing at 87.72 frames per second (FPS) with a low inference latency of 2.6ms and preprocessing time of 8.8ms. With only 1.48 million parameters, this lightweight design supports real-time processing, establishing a new benchmark for SWR inspection in industrial settings.

## I. INTRODUCTION

Steel wire ropes (SWRs) are critical components in various industrial applications, such as cranes, elevators, and bridges, due to their high strength and flexibility [1]. However, prolonged usage and harsh operating conditions can lead to various local flaws (LFs). At present, the nondestructive testing of SWR is mainly carried out through electromagnetic, visual, acoustic emission, eddy current, and so on. Among these methods, electromagnetic-based methods, particularly magnetic flux leakage (MFL), have made significant advancements in the detection of LFs [2], [3]. MFL is a widely used nondestructive testing method that detects defects by measuring perturbations in the magnetic field caused by the presence of LFs in the SWRs. The process involves magnetizing the SWR, and as the rope passes through a sensor array, any anomalies such as cracks or defects cause the magnetic flux to leak. These flux variations are then detected by the sensor array, providing valuable information regarding the location and size of the defects [5].

Recent advancements in deep learning have shown great potential in improving LF detection tasks by automatically extracting relevant features from raw data. One advantage of deep learning is that the neural network can automatically extract meaningful features and patterns without selecting features manually [6]. Current methods based on MFL detection typically use multi-channel signals collected from multi-channel sensors. Due to the variation of LF position and spatial diffusion characteristics of the LF signal, the features contained in different channels often differ. Therefore, choosing the proper form of input for a neural network can be a challenging step in designing a deep learning framework for multi-channel input.

At present, deep-learning based LF detection methods can generally be divided into two categories based on the input of neural networks: multi-dimensional methods and single-dimensional methods. Among the multi-dimensional methods, the most common approach is to use images and then apply computer vision techniques to detect defects. A typical method collects images of the surface of the rope and then uses object detection or image segmentation techniques for defect identification. Object detection algorithms, which possess the capability to accurately identify and localize target regions within images, are thus frequently employed in the detection of surface defects in SWRs. For example, Zhou et al. propose a YOLO-based intelligent algorithm for SWR for real-time surface damage detection in SWRs, leveraging computer vision and deep learning techniques to achieve high accuracy and efficiency in industrial applications [7]. However, in real-world applications, the surface images are susceptible to blur due to wire rope movement and equipment vibration. Therefore, Jiang et al. propose a method combining Generative Adversarial Networks for image deblurring and YOLOv7 for damage detection, effectively restoring blurred images and accurately identifying surface defects in SWR [8]. Such multi-dimensional methods based on SWR surface images are simple and direct, and the acquisition of surface images of SWR is relatively easy. However, it can only detect surface LFs and cannot detect internal LFs which also pose risks and cause severe damage if not detected.

Another multi-dimensional detection method involves interpolating the raw MFL signals to form MFL images. This method makes good use of the spatial characteristics of multi-channel signals and intuitively displays the distribution of MFL signals across different channels. Based on MFL images, Pan et al. utilized deep learning techniques, training a YOLOv5s object detection model to identify defective regions in MFL images, reaching over 96% accuracy [9]. Despite achieving excellent performance, training object detection models often require a significant amount of time.

Furthermore, the process of converting raw signals into MFL images necessitates interpolation, which increases the computational burden of data processing. Moreover, image-based models typically demand a large number of parameters. This undoubtedly complicates the deployment and real-time detection of edge devices, such as microcontrollers. Besides large parameters, interpolation of the signal does not add any extra determinant information to the original data but instead introduces new noise. For instance, when interpolating the strand noise into an image, it might exhibit characteristics similar to defects under high-speed, high-slope conditions [10]. This could potentially lead to some strand noise being misidentified as defects, resulting in false positives.

On the other hand, single-dimensional methods refer to utilizing one-dimensional data to diagnose LFs. One typical approach is compressing multi-channel signals into one-dimensional data by dimension-reduction methods and detecting LF by using compressed data. For example, Lu et al. utilize Principal Component Analysis (PCA), which is an unsupervised data compression technique, to compress multi-channel data into one dimension and use CNN to classify defects [11]. While this approach is simple and efficient, it also loses the spatial characteristics of the multi-channel MFL signal in other dimensions such as circumstantial dimension and radial dimension. Such signals in circumstantial dimension and radial contain important inter-channel information, which is also crucial in LF detection [12]. Another common one-dimensional approach is directly using single-channel signals collected by sensors to detect LFs. Liu and Chen utilize a one-dimensional convolutional neural network (CNN) to automatically extract local LF features in the time domain, achieving around 98% testing accuracy, significantly outperforming traditional handcrafted feature-based machine learning methods [13]. However, signals collected by a single channel are likely to be interfered with by the jittering or shaking during inspection [14]. Another drawback is that due to the discrepancy in the distribution of LFs under different LF and sensor positions, it is unknown whether a specific signal channel can detect LF signals during inspection, making it a more challenging utilization for a single-dimensional approach.

Based on the above analysis, both multi-dimensional and single-dimensional methods have their advantages and drawbacks, thus making us rethink the essence of detecting LF signals: detecting local abnormal distributions in multi-channel signal dimensions. While interpolation does not provide extra determinant information for LF inspection and dimension reduction causes loss in multi-dimension information, the collected multi-channel signal matrices themselves already contain sufficient multi-dimensional information, which is meant to be an ideal choice for balancing both computational complexity and precise information. Therefore, it drives us to think of using pre-processed multi-channel signal matrices as the input of neural networks, which not only preserve the multi-channel features but also reduce the computational load of the network compared with interpolation methods. Therefore, this article introduces a new LF detection paradigm based on the MFL multi-channel signal matrix. The contribution of this article can be summarized as follows:

1) We comprehensively analyze existing LF detection methods, identifying key limitations in both one-dimensional and image-based approaches, and propose a novel signal matrix-based paradigm that preserves multi-channel spatial relationships while avoiding interpolation artifacts and computational inefficiencies.

2) We develop a specialized CNN architecture designed for processing raw multi-channel MFL signal matrices, called SM-CNN. This architecture is tailored to handle the unique characteristics of signal matrices.

3) To optimize the performance of SM-CNN, we introduce specific components such as stripe pooling to manage asymmetric signal dimensions and symmetric padding to improve boundary defect detection. Additionally, we design a data augmentation strategy tailored for signal matrix inputs, which enhances the network's generalization capability.

The rest of this article is arranged as follows: Section II introduces the data acquisition process and corresponding processing techniques for data used in this article. Section III presents the methodology. Section IV illustrates and discusses the results of the proposed method and Section V concludes this article.

## II. MFL Inspection Equipment and Data Acquisition

### A. Data Acquisition

In this article, based on the MFL principle, the MFL sensor shown in Fig. 1 is designed and utilized to collect MFL signals.

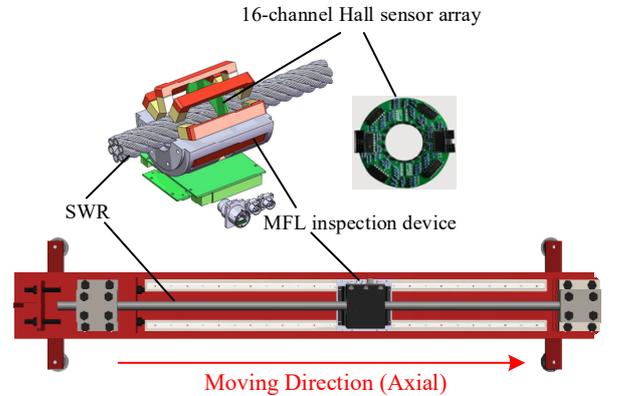

Fig. 1. The schematic of the MFL inspection device.

During the inspection process, the SWR to be tested is magnetized by magnets, thus formulating a magnetic circuit between the armatures made of irons and SWR. The material of magnets used in this equipment is NeFeB. Then, the Hall sensors array, which can transform magnetic signals into electrical signals, is used to detect MFL signals in the axial direction. To obtain circumstantial information on the SWR, 16 Hall sensors are uniformly arranged in a ring, which composes a uniform circular array. The inner diameter of the sensor is 40 mm. The sensor can be applied under two modes, which are portable mode and online mode respectively. This article uses portable mode in all experiments [15].

## B. Data Preprocessing

### 1) Smoothing and Detrending

After collecting the raw data, subsequent signal processing techniques are performed. As shown in Fig. 2(a), the raw multi-channel MFL data can be denoted as $x[m, n]$ for the $m$th sample value received in the $n$th channel, $m = 1, 2, \ldots, M$ and $n = 1, 2, \ldots, N$, where $M$ is sample number and $N$ is channel number. The raw MFL signals often contain high-frequency noise and baseline drift, which can obscure the defect-related features. Therefore, a low-pass filter is applied to the raw signals to remove high-frequency noise. This is achieved by convolving the signal with a smoothing kernel, such as a moving average filter or a Gaussian filter. The smoothed signal $\hat{x}[m,n]$ is obtained as follows:

$$\hat{x}[m,n] = \sum_{k=-K}^{K} w[k] \cdot x[m+k, n], \quad (1)$$

where $w[k]$ is the smoothing kernel, and $K$ is the kernel size, which is 5 in this article. This operation effectively reduces noise while preserving the underlying LF-related features.

After smoothing, baseline drift, often caused by sensor instability or environmental factors, is removed to ensure that the defect-related features are more prominent. A detrending operation is applied to the smoothed signal $\hat{x}[m,n]$ by subtracting a fitted trend line. The trend line is typically obtained using a polynomial fit or a linear fit. The detrended signal $\tilde{x}[m,n]$ is computed as:

$$\tilde{x}[m,n] = \hat{x}[m,n] - T[m,n], \quad (2)$$

where $T[m,n]$ represents the trend component of the signal. As shown in Fig. 2(b), this step ensures that the signal is centered around zero, making it easier to detect localized defects.

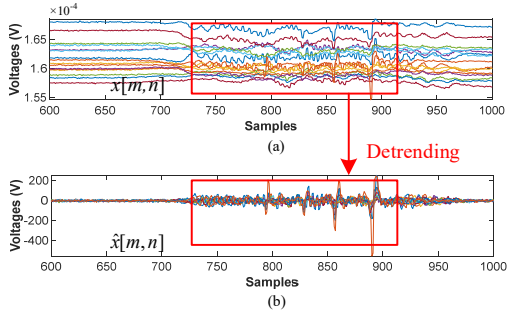

Fig. 2 The pre-processing process of MFL signals. (a) raw MFL signal (b) detrended MFL signals

### 2) Windowing and Normalization

After detrending, any residual baseline offset is removed by subtracting the mean value of the signal over a sliding window. This further enhances the visibility of defect-related features in the signal. The preprocessed signals are then divided into windows of 300 sampling points, resulting in a signal matrix of size 300×16. This signal matrix is then normalized and then prepared to be input to further processing.

## III. METHODOLOGY

### A. MFL Signal Analysis

To design an appropriate neural network structure for identifying LF in the MFL signal matrix, it is necessary to analyze the characteristics of the LF signal. As illustrated in Fig. 3(a), for single-channel MFL signal, there are mainly two components: LF signals and strand noise. LF signals manifest as pulses in the time domain. Strand noise, caused by variations in the lift-off distance between the sensor and the surface, exhibits periodic fluctuations resembling a sinusoidal waveform in the time domain. In the case of multi-channel MFL signal matrices, due to the spatial variations in the distribution of LFs and the differences between sensors in different channels, LF signals decay along the circumstantial direction as the distance between sensors and LF increases [16]. Therefore, only a subset of adjacent channels can detect LF signals. Among those channels that can detect LF signals, the MFL signals exhibit similar characteristics to those seen in single-channel detection. However, as the LF signal pulse diffuses circumferentially, it manifests as spatial pulses with two peaks and one valley on the signal matrix plane. To more clearly illustrate the spatial distribution of the LF signal, we interpolated the original 16 channels into 200 channels, resulting in a spatial distribution image, as shown in Fig. 3(b). In real-world scenarios, the position of the defect is typically unknown, which means the exact timing of LF signal appearances in the time domain is also unknown.

Based on the analysis above, the characteristics of LF signals in the two-dimensional signal matrix can be summarized as a form of localized anomaly that is randomly distributed on the signal matrix. To extract local features of LF in a signal matrix, 2D-CNN, which is particularly effective at capturing local features, is chosen to be the backbone.

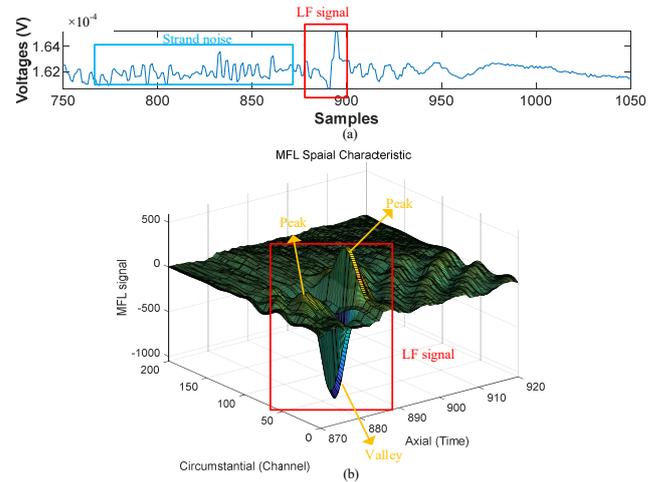

Fig. 3. The example of MFL signal (a) single-channel MFL signal (b) MFL signal spatial characteristic.

### B. CNN Optimization Strategies for Signal Matrix Input

Traditional CNN architectures are primarily designed for two-dimensional image data. However, when applying CNNs to classify or recognize signal matrices, it is necessary to adapt the network to the specific characteristics of the signal matrices. Unlike images, which typically have two spatial dimensions with similar sizes, signal matrices often exhibit significant asymmetry in their axial and circumferential dimensions. Specifically, the axial dimension (corresponding to the time dimension) is much larger than the circumferential dimension (corresponding to the channel dimension). To address this, some optimization techniques are proposed.

*1) Stripe Pooling*

The signal input matrix in our application has a size of 16×300, where 16 represents the number of sensor channels and 300 corresponds to the time dimension. Traditional CNN architectures typically utilize square pooling operations, where the pooling kernel is of equal size along both dimensions. However, for signal matrices with highly asymmetric input dimensions, applying square pooling kernels results in excessive compression along the channel dimension after just a few pooling layers. This rapid reduction in the channel dimension limits the network's depth, hindering its ability to capture higher-level features.

To address this issue, we propose the use of stripe pooling, where the pooling operation is specifically designed to account for the dimensional imbalance. In the early layers of the network, we apply a 2×1 rectangular pooling kernel, which only compresses the time dimension (300) and preserves the channel dimension (16). This reduces the compression along the channel axis and allows the network to maintain a deeper structure, facilitating the extraction of richer features. Moreover, due to its long and narrow kernel shape, it can easily establish long-range dependencies between discretely distributed regions and effectively encode strip-shaped areas [17].

The low information density in the circumferential direction introduces another issue. As the sensor pose changes and LFs are distributed at different positions in the SWR, the distribution of defect signals across signal channels may vary, potentially leading to a specific scenario where the LF signal distributes at the upper end and lower end of the MFL signal matrix. To intuitively demonstrate this variation, an interpolation method is employed here to visualize the signal matrix.

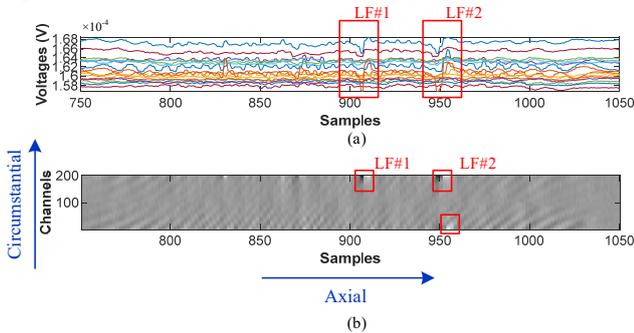

Fig. 4. A sample MFL data with two LFs distributed at the two ends of a signal matrix (a) raw MFL signals (b) MFL image interpolated from a signal matrix.

As shown in Fig. 4, two LFs are located at the ends of the signal matrix. Through interpolation, the spatial characteristics of these two LFs are magnified and can be visualized. For image-based LF detection methods, increasing the circumferential information density via interpolation amplifies the localized LF features distributed at the ends of the signal matrix, thereby generating visually identifiable LF characteristics on a macroscopic scale. Consequently, for image-based detection methods, techniques such as template matching or object detection can still extract these features, enabling the detection of such LFs. However, for the signal matrix as an input modality, the low circumferential information density causes the LF signals distributed at the matrix end to often concentrate in the topmost and bottommost channels. Direct application of convolutional operations fails to effectively extract these features, and the network is unable to detect such specially distributed LFs. To mitigate this issue, we use symmetric padding for the signal matrix, which is shown in Fig. 5.

For the channels at the ends of the matrix, we copy the first channel and append it to the bottom, and similarly, the last channel is copied and appended to the top. This approach helps avoid the problem of low circumferential information density, ensuring that the convolution layers can still capture important features near the boundaries.

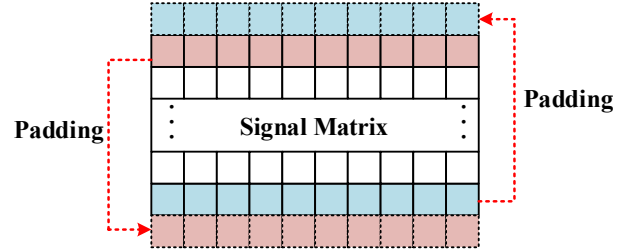

Fig. 5. The schematic of symmetrical padding

*B. Overview of SM-CNN*

Based on the characteristics of a signal matrix, a CNN specialized for signal matrix (SM-CNN) is proposed. The detailed structure is illustrated in Table I.

TABLE I. THE STRUCTURE OF THE PROPOSED SM-CNN

| Layer | Size | Channel | Activation | Tensor shape |
|---|---|---|---|---|
| Input | - | - | - | (300,16,1) |
| Symmetric Padding | - | - | - | (300,18,1) |
| Conv2D | (2,2) | 16 | ReLU | (299,17,16) |
| MaxPooling2D | (2,1) | - | - | (149,17,16) |
| Conv2D | (2,2) | 32 | ReLU | (148,16,32) |
| MaxPooling2D | (2,1) | - | - | (74,16,32) |
| Conv2D | (2,2) | 64 | ReLU | (73,15,64) |
| MaxPooling2D | (2,1) | - | - | (36,15,64) |
| Conv2D | (2,2) | 128 | ReLU | (35,14,128) |
| MaxPooling2D | (2,1) | - | - | (17,14,128) |
| Conv2D | (2,2) | 256 | ReLU | (16,13,256) |
| MaxPooling2D | (2,2) | - | - | (8,6,256) |
| Flatten | - | - | - | 12288 |
| Dense | 128 | - | ReLU | 128 |
| Output | 2 | - | Softmax | 2 |

The input to the model is a 2D signal matrix with a shape of (300, 16, 1), where 300 represents the width (temporal or spatial dimension) and 16 represents the height (circumferential dimension). To address the issue of low circumferential information density and ensure effective feature extraction near the boundaries, symmetric padding is applied to the input matrix. Specifically, the first channel (top row) is copied and appended to the bottom, and the last channel (bottom row) is copied and appended to the top, increasing the height dimension to 18.

The model employs a series of convolutional and pooling layers to extract features. In the early stages of the network, stripe pooling with a pool size of (2, 1) is used to avoid the problem of the network being unable to deepen due to pooling in the signal channel dimension. This approach preserves the circumferential structure while reducing the width dimension. Additionally, small convolutional kernels of size (2, 2) are utilized to fully capture the relationships between channels. The model architecture consists of multiple convolutional layers with increasing filter sizes (16, 32, 64, 128, 256), each followed by a ReLU activation function and a pooling layer. The convolutional layers focus on extracting features along the width dimension, while the pooling layers progressively reduce the spatial dimensions.

After the convolutional and pooling operations, the feature maps are flattened into a one-dimensional vector and passed through a fully connected layer with 128 neurons and ReLU activation. Finally, the output layer uses 2 neurons with Softmax activation to perform binary classification.

### C. Implementation Details

The SM-CNN model is implemented using the Keras framework and Python 3.8, and all training and testing processes are performed on a laptop with an AMD Ryzen 7 7840HS CPU and 32G RAM. During the training process, the learning rate is 0.0001, and the batch size is 32. The Adam optimizer is used for model training. The model is trained by 10 epochs. Binary cross-entropy loss is chosen as the loss function, which is appropriate for classification tasks. The categorical cross-entropy loss can be calculated as:

$$L = -\left( y \log(p) + (1-y) \log(p) \right), \quad (3)$$

where $y$ is the true label (either defect or normal), and $p$ is the predicted probability of positive class (defect).

To comprehensively evaluate the performance of the proposed methods, several metrics for classification tasks are used, which are accuracy, precision, recall, and F1-score. The dataset contains 196 LF signal samples and 202 normal signal samples. The ratio of train set to test is 7:3.

## IV. RESULTS AND DISCUSSIONS

### A. Experimental Results

After training the modal, the loss curve and accuracy curve with iterations are shown in Fig. 6. The results of the proposed method are shown in Table II. To demonstrate the superiority of the proposed method based on the signal matrix, we further compare it with the state-of-the-art methods and three baseline methods. The first baseline method is to use PCA to reduce the dimension of the multi-channel signal matrix and use a constant threshold for detecting LF areas. The second method is based on a 1D-CNN which uses a single channel as the input. The results of 16 channels are averaged as the final results. The third method is shaking noise elimination (SNE), which uses a constant threshold to process signals after filtering out shaking noise.

From Table II, the proposed SM-CNN demonstrates an accuracy of 98.74%, precision of 99.35%, recall of 97.85%, and F1-score of 98.60%, outperforming existing methods and baseline methods. In terms of the 1D-CNN method which uses a single-channel signal as the input, it has an overall F1-score of 84.49%. However, its recall (75.75%) is significantly lower than the proposed method (97.85%). Such a significant performance gap can be attributed to the fact that many signal channels do not exhibit clear signal features, making it difficult for the model to correctly identify these defects. As a result, numerous defects were not correctly detected, ultimately leading to the lower recall rate observed in the 1D-CNN method. This highlights the importance of multi-channel inputs, which can capture both time-wise and channel-wise LF characteristics, improving recall and detection accuracy overall.

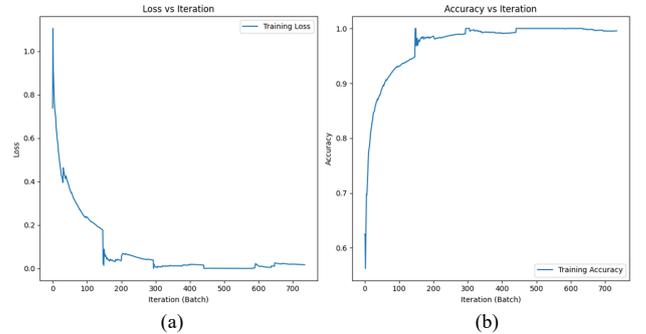

Fig. 6. The loss and accuracy curves during the training process. (a) training loss curve. (b) training accuracy curve.

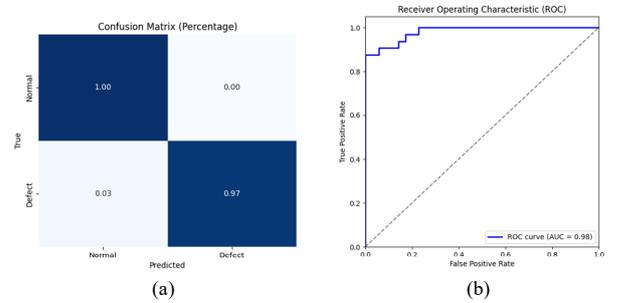

Fig. 7. The classification results (a) confusion matrix. (b) ROC curve

To further demonstrate the effectiveness of signal matrix as the model input, we further compared with image-based SOTA methods including target-oriented-feature (TFO) method, adaptive spatial sampling rate TFO (AS-TFO), and circular noise suppression (CNS). Compared with these image-based methods, the proposed SM-CNN also demonstrates superior performance. While image-based methods have relatively high recall, their precision is lower (86.07% for TFO and 90.32% for CNS). This is due to that image methods rely on interpolation to generate MFL images from raw signals. This interpolation magnifies the effects of noise, causing false negatives and lower precision.

Meanwhile, compared with data augmentation techniques such as random rotation, and color jittering, the proposed data augmentation technique generates high-quality samples with very low computational costs, further improving the generalization and overall performance. This also indicates the great potential of this signal matrix-based detection paradigm.

In addition to performance comparison, we also give the confusion matrix and Receiver Operating Characteristic (ROC) curve of the proposed SM-CNN results, which are shown in Fig. 7.

### B. Evaluation of Computational Efficiency

To evaluate the computational efficiency superiority of the proposed signal matrix detection framework and the TFF-CNN model, we further compared their computational speed with state-of-the-art methods. The comparison metrics mainly include the preprocessing time required from the raw signal to the network input, the model inference time, the number of model parameters, frame per second (FPS), and the giga floating point operations per second (GFLOPs). These metrics provide a comprehensive evaluation of computing time and model complexity. The statistical results are presented in Table III.

TABLE II. THE PERFORMANCE COMPARISONS WITH BASELINE METHODS

| Method | Input form | Accuracy (%) | Precision (%) | Recall (%) | F1-score(%) |
|---|---|---|---|---|---|
| PCA + Constant threshold | Signal | 81.25 | 83.58 | 75.22 | 78.77 |
| Single-channel 1D-CNN | Signal | 85.00 | 96.86 | 75.75 | 84.49 |
| SNE [18] | Signal | - | 29.89 | 92.86 | 45.22 |
| TFO [19] | Image | - | 86.07 | 94.68 | 94.68 |
| AS-TFO [10] | Image | - | 95.99 | 98.36 | 97.16 |
| CNS [20] | Image | - | 90.32 | 93.33 | 91.80 |
| YOLOv5s [9] | Image | 96.31 | 96.40 | 96.40 | 96.40 |
| SM-CNN | Signal Matrix | 96.87 | 95.70 | 97.87 | 96.62 |
| SM-CNN+ Augmentation | **Signal Matrix** | **98.74** | **99.35** | **97.85** | **98.60** |

TABLE III. THE EVALUATION OF COMPUTATIONAL EFFICIENCY WITH SOTA METHODS

| Method | Modality | Preprocessing Time (ms) | Inference Time (ms) | FPS | Params | GFLOPs |
|---|---|---|---|---|---|---|
| TFO [19] | Image | 19.6 | 290.6 | 3.23 | - | - |
| AS-TFO [10] | Image | 19.6 | 300.5 | 3.12 | - | - |
| YOLOv5s [9] | Image | 19.6 | 50 | 14.37 | 7.24M | 16.5 |
| SM-CNN | Signal Matrix | **8.8** | **2.6** | **87.72** | **1.48M** | **0.166** |

From Table III, the SM-CNN model exhibits a preprocessing time of 8.8 ms, slightly lower than image-based methods (19.6 ms) due to the cancellation of the interpolation step. Moreover, SM-CNN achieves a significantly faster inference time of 2.6 ms, outperforming TFO (290.6 ms), AS-TFO (300.5 ms), and YOLOv5s (50 ms). Additionally, SM-CNN has only 1.48 million parameters and 0.166 GFLOPs, making it more lightweight and computationally efficient than YOLOv5s (7.24M parameters, 16.5 GFLOPs). These results demonstrate that SM-CNN and the proposed signal-matrix detection paradigm strike an excellent balance between preprocessing overhead and computational efficiency, making it highly suitable for real-time industrial applications.

## V. CONCLUSION

In this article, we present a novel signal matrix-based LF detection paradigm and design a CNN framework that achieves state-of-the-art performance by directly processing raw multi-channel MFL signals without interpolation. The proposed SM-CNN architecture addresses key challenges in MFL signal analysis through innovative techniques including stripe pooling to maintain channel-wise features and symmetric padding to enhance boundary defect detection. Experimental results demonstrate exceptional accuracy (98.74%) and recall (97.85%) while maintaining real-time processing capability (87.72 FPS) with a lightweight model (1.48M parameters). The framework's computational efficiency (2.6 ms inference latency) and robust performance establish it as a practical solution for industrial SWR inspection systems. These results provide strong evidence of the effectiveness of the proposed detection paradigm. Future work will focus on extending our model and dataset for more flaw types and shaking noise classification and deployment on embedded devices.

## ACKNOWLEDGMENT

This work was supported by the Sichuan Science and Technology Program under Grants 2024JDHJ0057 and MZGC20240050.